\documentclass[conference]{IEEEtran}
\IEEEoverridecommandlockouts
\usepackage{cite}
\usepackage{amsmath,amssymb,amsfonts}
\usepackage{algorithmic}
\usepackage{graphicx}
\usepackage{textcomp}
\usepackage{xcolor}
\def\BibTeX{{\rm B\kern-.05em{\sc i\kern-.025em b}\kern-.08em
    T\kern-.1667em\lower.7ex\hbox{E}\kern-.125emX}}
\begin{document}

\title{Accelerating Cardiac MRI Reconstruction with CMRatt: An Attention-Driven Approach
}

\author{\IEEEauthorblockN{Anam Hashmi}
\IEEEauthorblockA{\textit{ML-Labs} \\
\textit{Dublin City University}\\
Dublin, Ireland \\
anam.hashmi2@mail.dcu.ie}
\and
\IEEEauthorblockN{Julia Dietlmeier}
\IEEEauthorblockA{\textit{Insight SFI Research Centre for Data Analytics} \\
\textit{Dublin City University}\\
Dublin, Ireland \\
julia.dietlmeier@insight-centre.org}
\and
\IEEEauthorblockN{Kathleen M. Curran}
\IEEEauthorblockA{\textit{School of Medicine} \\
\textit{University College Dublin}\\
Dublin, Ireland \\
kathleen.curran@ucd.ie}
\and
\IEEEauthorblockN{Noel E. O'Connor}
\IEEEauthorblockA{\textit{Insight SFI Research Centre for Data Analytics} \\
\textit{Dublin City University}\\
Dublin, Ireland \\
noel.oconnor@insight-centre.org}

}

\maketitle

\begin{abstract}
Cine cardiac magnetic resonance (CMR) imaging is recognised as the benchmark modality for the comprehensive assessment of cardiac function. Nevertheless, the acquisition process of cine CMR is considered as an impediment due to its prolonged scanning time. One commonly used strategy to expedite the acquisition process is through k-space undersampling, though it comes with a  drawback of introducing aliasing effects in the reconstructed image. Lately, deep learning-based methods have shown remarkable results over traditional approaches in rapidly achieving precise CMR reconstructed images. This study aims to explore the untapped potential of attention mechanisms incorporated with a deep learning model within the context of the CMR reconstruction problem. We are motivated by the fact that attention has proven beneficial in downstream tasks such as image classification and segmentation, but has not been systematically analysed in the context of CMR reconstruction. Our primary goal is to identify the strengths and potential limitations of attention algorithms when integrated with a convolutional backbone model such as a U-Net. To achieve this, we benchmark different state-of-the-art spatial and channel attention mechanisms on the CMRxRecon dataset and quantitatively evaluate the quality of reconstruction using objective metrics.  Furthermore, inspired by the best performing attention mechanism, we propose a new, simple yet effective, attention pipeline specifically optimised for the task of cardiac image reconstruction that outperforms other state-of-the-art attention methods. The layer and model code will be made publicly available.

\end{abstract}


\section{Introduction}

Cardiac magnetic resonance (CMR) imaging is used as a fundamental clinical tool for the comprehensive evaluation of cardiovascular diseases. CMR imaging is widely acknowledged for its non-invasiveness and its ability to generate high-resolution images without resorting to ionizing radiation \cite{larose2007cardiovascular}. However, there is a major drawback associated with the MRI acquisition process which is their prolonged scanning time. To address this problem i.e. to reduce scanning times, undersampling in k-space (Fourier domain) is often used, though it leads to aliasing artifacts in the reconstructed image. 

However, reconstructing CMR images from undersampled k-space poses a significant challenge with traditional approaches like Compressed Sensing (CS) \cite{10251064} being computationally time consuming. In contrast, deep learning (DL) based models have marked a transformative shift in accelerated CMR reconstruction and have shown superior performance to traditional methods \cite{yiasemis2023deep}. Therefore, the construction of a robust deep Convolutional Neural Network (CNN) is critical for achieving enhanced outcomes. 
To achieve this, numerous lightweight plug-and-play attention modules have been designed to efficiently capture global context information and suppress irrelevant features and thus enhance the representational properties of CNNs. These attention modules are designed to seamlessly integrate into a CNN architecture, ensuring minimal computational overhead. 

In general, attention mechanisms, often referred to as feature recalibration modules, inspired by human visual perception, represent a valuable addition to deep learning models as they excel in amplifying the crucial image features while attenuating the irrelevant ones \cite{10251064}. However, navigating the existing research landscape concerning the unified performance analysis of attention in medical imaging is a challenging task. The common trend in the majority of the state-of-the-art attention studies present performance gains in terms of classification accuracy metrics or using top-1 and top-5 error rates, underscoring that these modules were specifically designed for classification, object detection, and segmentation tasks \cite{hu2018squeeze, ruan2021gaussian}. Moreover, the datasets commonly used in these studies are typically not from the medical imaging domain. Instances of such datasets include ImageNet for classification, COCO for object detection and Kinetics for video classification \cite{yang2020gated}.

Recently, a new large-scale public CMRxRecon dataset \cite{wang2023cmrxrecon} including multi-contrast, multi-view and multi-slice CMR imaging data from 300 subjects has been introduced to the research community at the CMRxRecon reconstruction challenge hosted at MICCAI 2023. The top-performing teams have developed pipelines integrating a combination of physics-based, deep learning-based and attention-based approaches \cite{xin2023fill, yiasemis2023deep}. Some studies employed modified 2D U-Net networks \cite{ronneberger2015u} with attention mechanisms \cite{xin2023fill, dietlmeier2023cardiac}. However, the addition of attention mechanisms lacks supporting performance analysis and we argue that integrating task-specific attention modules could further improve the results. Thereby, we conclude that there is a timely need for a performance analysis study on attention methods in the image reconstruction task, specifically focusing on medical MRI datasets and considering objective image reconstruction quality metrics such as Peak Signal to Noise Ratio (PSNR), Mean Square Error (MSE) and Structural Similarity Index Measure (SSIM) \cite{mason2019comparison}.

This observation has motivated us to conduct the first systematic study on the addition of attention mechanisms with a well-known deep learning model for the CMR reconstruction problem. To the best of our knowledge,this is the first study comprehensively exploring different state-of-the-art attention modules in the context of CMR reconstruction. To sum up, the main contributions of our research are: 
\begin{itemize}
    \item We conduct extensive experiments to benchmark different state-of-the-art attention blocks by integrating them into a U-Net based CMR reconstruction model on the CMRxRecon dataset.
    \item We determine the attention modules that most significantly enhance the performance of the reconstruction task by ranking them based on the SSIM metric.
    \item Inspired by the best-performing algorithm, we propose a new, simple yet effective, attention pipeline for the CMR reconstruction task. Our proposed pipeline outperforms the state-of-the-art attention blocks.
\end{itemize}

\section{Related Research}
Recently, several attention modules 
have been proposed to capture rich global feature relationships in CNNs. These attention modules can be broadly categorised into channel attention, spatial attention, temporal attention and branch attention \cite{guo2022attention}. In this work, we focus on state-of-the-art (SOTA) channel, spatial and hybrid attention mechanisms, integrating both channel and spatial components. One representative module is  Squeeze-and-Excitation (SE) attention \cite{hu2018squeeze}, designed to capture cross-channel relationships and recalibrates feature responses on a channel-wise basis. Following SE attention, CBAM \cite{woo2018cbam} and BAM \cite{park2018bam} were proposed to refine feature maps by integrating both channel and spatial attentions. Another attention network sharing similar structure to SE called Global-Context Networks (GC-Net) \cite{cao2019gcnet} was proposed. GC-Net aimed at capturing long-range dependencies and global context features more effectively by employing non-local blocks. Following the advance of style transfer with CNNs, a Style-based Recalibration Module (SRM) \cite{lee2019srm} that leverages style information to reweight the feature maps was proposed. To reduce the parameters introduced by the attention mechanisms, a parameter-free Gaussian Context Transformer (GCT) \cite{ruan2021gaussian} attention module was introduced, utilizing a Gaussian function to capture global contextual feature excitation. A recent and computationally inexpensive module, Attention Bias (AB) \cite{klomp2023performance}, was proposed, utilizing the Hadamard product for recalibrating the feature maps. In contrast to the above mentioned channel and spatial attention modules, a parameter free block, SimAM \cite{yang2021simam} was designed to calculate 3-D attention weights in a feature map. This attention block is based on the property of spatial suppression and uses an energy function for calculating feature weights. A trending attention mechanism is Self-attention \cite{vaswani2017attention}, also known as intra-attention, which effectively captures long-range dependencies. The performance of transformers relies significantly on the integration of stacked self-attention blocks. However, all the above referenced papers conducted experiments on their attention blocks in the context of classification, segmentation and object detection tasks. Only a very limited number of studies have reported incorporation of attention blocks in their U-Net-based networks for the task of cardiac MRI reconstruction from undersampled k-space. Some of the studies to do so include \cite{xin2023fill, huang2019mri, dietlmeier2023cardiac}.

\section{Network Architecture}
This paper is an extension of our previous study \cite{dietlmeier2023cardiac} for CMR reconstruction with an objective to delve deeper into the integration and performance analysis of attention mechanisms and further it. Therefore, a vanilla U-Net architecture \cite{ronneberger2015u} serves as the baseline model for our investigation. 

In this study, we worked with 2D slices and built a 2D U-Net with four encoder and four decoder blocks connected through a bridge section. Each encoder subnetwork of U-Net consisted of two $3\times3$ convolutions followed by a batch normalization layer and a ReLU (Rectified Linear Unit) activation function, referred to as the conv\textunderscore block. The ReLU output functioned as the skip connection for the corresponding decoder block. To regularize our model, we also added dropout layers with dropout probability of 0.25 in each of four encoder blocks which is followed by a $2\times2$ max pooling layer. In the initial downsampling stage, 32 feature maps were employed. The decoder block mirrors the encoder block in reverse and consisted of the transposed convolution layer, concatenated with the corresponding skip connection from the encoder block, and finally followed by the above mentioned conv\textunderscore block. 

\begin{table*}[htbp]
\centering
\caption{ Our experimental results of benchmarking state-of-the-art attention mechanisms on the CMRxRecon dataset. 
Methods are ranked according to the SSIM metric. Computational overhead is given in terms of model parameters.} 

\begin{tabular}{llllll}

\hline
\hline
\textbf{Method} & \textbf{ Parameters} & \textbf{Computational overhead} &\textbf{PSNR} $\uparrow$ & \textbf{ MSE} $\downarrow$ & \textbf{SSIM} $\uparrow$\\
\hline
Baseline (without attention)      & see Section III    & 0         & 36.20684 & 0.00028783                                            & 0.92451  \\
\hline
SimAM  \cite{yang2021simam}     & lambda=1e-4   & 0            & 37.04925 & 0.00025836 & 0.94432  \\
SE \cite{hu2018squeeze}        & reduction=16   & 119,936             & 37.28633 & 0.00023553                                           & 0.94297  \\
CAB \cite{xin2023fill}        & reduction=4, kernel size=3 & 17,750,529               & 36.75172 & 0.00030422                                           & 0.94206  \\
GCT \cite{ruan2021gaussian}    & c=2         &0                & 36.58741 & 0.00026953                                           & 0.94088  \\
CBAM \cite{woo2018cbam}        & reduction=16, kernel size=7 &122,092 & 37.62119 & 0.00023203                                           & 0.93845  \\

AB \cite{klomp2023performance} & None        &3,424                & 34.36335 & 0.00042996                                           & 0.93570  \\
GC-Net \cite{cao2019gcnet}     & reduction=16     &127,448           & 36.67425 & 0.00026559                                           & 0.90888  \\
BAM \cite{park2018bam}         & reduction=16  & 259,632              & 34.52592 & 0.00039782                                           & 0.90293  \\
SRM \cite{lee2019srm}          & None      & 13,696                  & 33.27101 & 0.00053154                                           & 0.90292\\
\hline
\end{tabular}
\end{table*}


\subsection{Comparative Analysis of State-of-the-art Attention Mechanisms: A Benchmarking Approach}
We incorporated various attention modules into our baseline U-Net model and conducted a comprehensive comparative analysis of their performances. The findings from our benchmarking study, conducted on nine state-of-the-art attention modules using the CMRxRecon dataset, are presented in Table 1. The experimental results highlight that the SimAM attention module \cite{yang2021simam} performed the best for the CMR reconstruction task, based on the SSIM metric, which is considered more robust than PSNR and MSE \cite{sara2019image}. The SimAM attention module integrated with the baseline model demonstrated superior performance to the baseline alone.


\begin{figure*}[t!]
    \includegraphics[width=\linewidth]{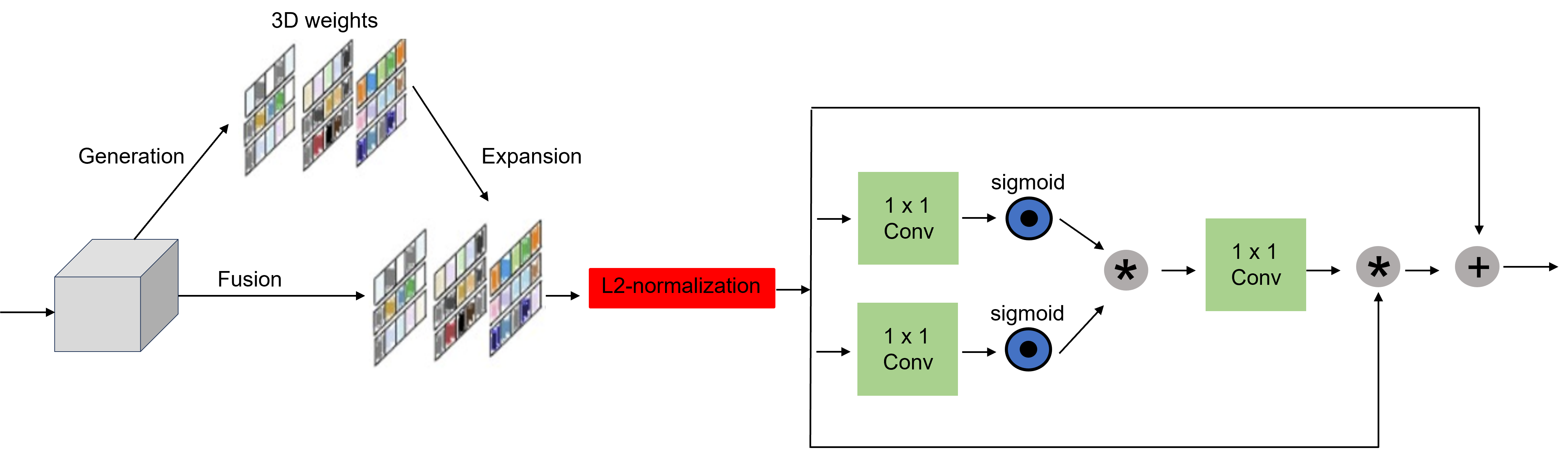}
    \caption{Proposed CMRatt attention framework, comprising SimAM attention block, L2-normalization layer, and a modified Hadamard attention.}
    \label{fig:1}
\end{figure*}

\subsection{Proposed Attention Pipeline}
Our experimental results demonstrate the superior performance of the SimAM \cite{yang2021simam}, making it the ideal choice as a key component in our proposed pipeline \textit{CMRatt} illustrated in Fig.1. In a nutshell, our design novelty is to combine SimAM with a L2 normalization layer and a modified Hadamard block \cite{jin2023simplified}. SimAM is a very lightweight attention module based on a neuroscience theory of spatial suppression \cite{webb2005early}.  
The SimAM block effectively generates 3D attention weights by optimizing an energy function to determine the importance of each neuron. 
To estimate the importance of each neuron, linear separability between one target neuron and other neurons is calculated as:
\begin{equation}
e_t^* = \frac{4(\hat{\sigma}^2 + \lambda)}{(t - \hat{\mu})^2 + 2\hat{\sigma}^2 + 2\lambda}
\label{eq:(1)}
\end{equation}
In Equation (1), \( t \) represents the target neuron, \( \hat{\sigma}^2 \) represents the variance of other neurons except \( t \), \( \hat{\mu} \) represents the mean of other neurons except \( t \), and \( \lambda \) is the coefficient. According to this equation, neurons with lower energy values show greater separability from other neurons, signifying their higher importance \cite{cai2023repvgg}. Sometimes, the values of the energy function can be too large and to restrict those, a sigmoid function is used. The final feature refinement equation is:
\begin{equation}
\tilde{X} = \text{sigmoid}\left(\frac{1}{E}\right) \odot X
\label{eq:your_label}
\end{equation}
In Equation (2), \( E \) represents the sum of energy functions \( e_t^* \) across channel and spatial dimensions.

Normalization layers have become a fundamental component of modern deep neural networks \cite{shao2020normalization}. An important addition to our attention pipeline is the L2 normalization layer that serves the purpose of promoting stability in the model's learning process by stabilizing the gradients. 
Moreover, we integrate a different form of self-attention to our proposed pipeline by taking inspirations from Hadamard attention \cite{jin2023simplified} and Criss-Cross attention \cite{huang2019ccnet}. 
The aim of our proposed pipeline is to enhance the model's ability to capture nuanced and subtle patterns inherent in cardiac MRI images, thus contributing to improved reconstructions. Therefore, incorporating a self-attention block becomes essential to capture rich contextual relationships. However, self attention blocks are computationally expensive with both time and space requirements scaling quadratically with the length of the input \cite{keles2023computational}. Due to memory limitations, we could not employ a transfomer self-attention block \cite{vaswani2017attention} into our pipeline. Instead, we opted for a simpler Hadamard attention block with minute changes. This block consists of two $1\times1$ convolutional kernels to enhance the relevant features and also to reduce the number of channels, thus aiding in reducing the overall computational complexity of the network. Following this, two Sigmoid activation functions are applied owing to their better expressive ability \cite{dai2021attention}, succeeded by the Hadamard product between the two. Subsequently, a $1\times1$ convolution is applied and the resultant is multiplied by the input. 
 The final output is obtained by adding this to the connection from the input.

The Proposed CMRatt is integrated after every convolutional layer and within the skip connections in the baseline U-Net model.

\section{Methodology}
The released CMRxRecon dataset \cite{wang2023cmrxrecon} contained a total of 300 healthy volunteers from a single center (raw k-space data obtained on three Tesla magnets). The dataset is divided into 120 volunteers for training , 60 for validation, and 120 for test data. 
The training set for cine reconstruction included fully sampled k-space single-coil data and undersampled k-space data with acceleration factors ×4, ×8 and ×10. We performed the experiments on the training set of CMRxRecon using only the long-axis fully sampled (Ground Truth) and undersampled single-coil data with the acceleration factor of ×10 as these contained the strongest aliasing artifacts.

\subsection{Preprocessing}
To preprocess the training set which has been provided in the frequency domain as k-space data, an Inverse Fast Fourier Transform (IFFT) was applied to obtain the images in the spatial domain. Subsequently, we further processed the single-channel 2D grayscale images, and rescaled them by the maximum image intensity. Additionally, a linear scaling transform was applied to ensure that the pixel intensities fall in the range [0, 1]. 
We uniformly resized all IFFT-reconstructed images to dimensions of 256 × 256 pixels to serve as an input to the U-Net model.

\subsection{Implementation Details}
The processing pipeline was implemented in Python 3.11.5 and the PyTorch framework. 
All experiments were conducted on a GTX4090 GPU with 24 GB of memory. For model training, we used an AdamW optimizer 
with a learning rate of 0.001. 
The training was done with a batch size of 2, and all models were trained for 300 epochs to ensure convergence. The Mean Square Error (MSE) was used as the main optimization objective and no data augmentation was used in the training phase.

\subsection{Evaluation Metrics}

Mean Square Error (MSE) is the most widely used image quality assessment metric with better values closer to zero. The MSE between two images $\hat{y}$ and $y$ are defined as follows:
\begin{equation}
\label{eq:(3)}
\text{MSE} = \frac{1}{MN} \sum_{m=1}^{M} \sum_{n=1}^{N} [\hat{y}(n, m) - y(n, m)]^2
\end{equation}

PSNR is defined as the ratio between the maximum possible signal power and the power of the distorting noise. This ratio between two images is computed in the decibel form as \cite{sara2019image}:

\begin{equation}
\label{eq:(4)}
\text{PSNR} = 10 \cdot \log_{10}\left(\frac{\text{peakval}^2}{\text{MSE}}\right)
\end{equation}where \text{peakval} is the maximum possible intensity value.

The Structural Similarity Index Measure (SSIM) is a perception-based metric that captures the mutual dependencies among adjacent pixels to assess the similarity of two images, such as brightness, contrast and structural properties: 

\begin{equation}
\label{eq:(5)}
\text{SSIM}(y, \hat{y}) = \frac{(2\mu_y\mu_{\hat{y}} + c_1)(2\sigma_{y\hat{y}} + c_2)}{(\mu_y^2 + \mu_{\hat{y}}^2 + c_1)(\sigma_y^2 + \sigma_{\hat{y}}^2 + c_2)}
\end{equation}where \( \mu_y \) and \( \mu_{\hat{y}} \) represent the mean values of the model output \( \hat{y} \) and the target output y, \( \sigma_y \) and \( \sigma_{\hat{y}} \) denote the corresponding pixel variance values and $\sigma_{y\hat{y}}$ is the covariance. 


\subsection{Experimental results}
To allow fast experimentation, we do not use all training data provided but construct the training subset S1 (single coil) as follows: we first process all timeframes and slices in the supplied cine lax.mat files \cite{dietlmeier2023cardiac}. We further randomly subsample the full training set and select 1,000 LAX images for training and 200 LAX images for testing. Our experimental benchmarking results were shown in Table 1. For all attention methods benchmarked, we used the standard value of the parameters from their papers.

\begin{figure*}[t!]
    \includegraphics[width=\linewidth]{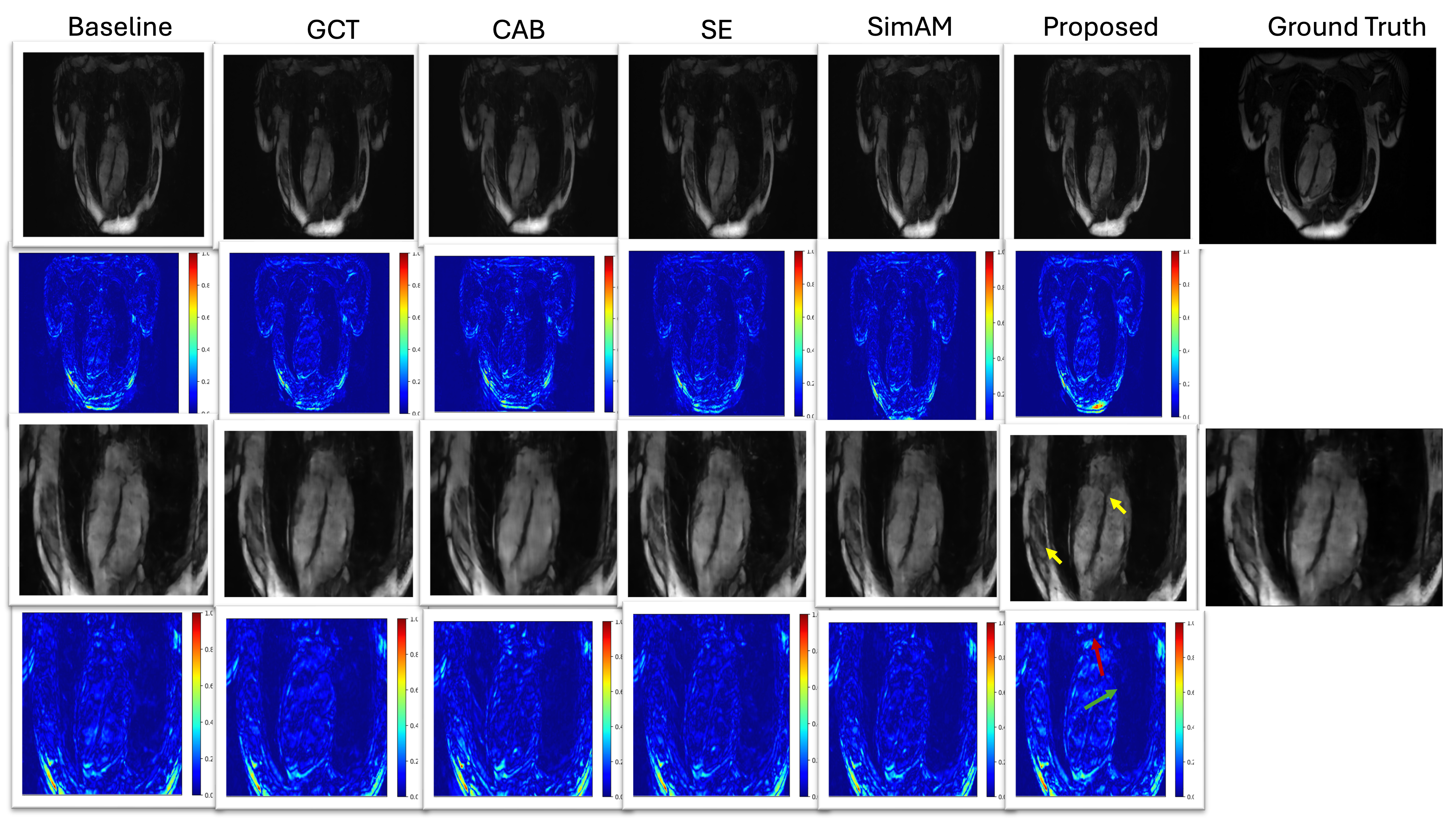}
    \caption{The reconstruction results and normalized absolute error maps of different methods of Patient 002 with acceleration factor $\times10$. The last two rows show the zoomed area. Proposed method can recover fine details on reconstructed images as marked by arrows. }
    \label{fig:2}
\end{figure*}

Without bells and whistles, \textbf{our proposed CMRatt mechanism significantly outperformed the baseline and the SimAM in all three metrics} and we achieved PSNR = 39.95486, MSE = 0.0001798 and SSIM = 0.96534. Moreover, in order to obtain further improvements in the reconstruction pipeline, consideration could be given to increasing the number of encoder-decoder blocks and the input resolution (e.g. $512\times512$) in the U-Net model. Our CMRatt, with a computational overhead of 969,870 model parameters, \textbf{demonstrated competitiveness and outperformed the other state-of-the-art attention blocks including the Channel Attention Block (CAB) which has a substantial computational overhead of 17,750,529 parameters, used by the winners of CMRxRecon challenge \cite{xin2023fill} in all the three quantitative metrics.} The qualitative results are shown in Fig.2. The qualitative comparisons demonstrate that our approach is capable of capturing fine details in the reconstructed images. 

\section{Conclusion}
This study systematically explored and benchmarked state-of-the-art attention mechanisms in the context of CMR reconstruction. 
Moreover, this study investigated the computational overheads associated with each attention module, thus addressing concerns related to memory limitations. Furthermore, we proposed a new CMRatt attention pipeline which outperformed the baseline, SimAM and the most recent CAB attention mechanisms. In conclusion, our study emphasized the potential of attention mechanisms to significantly improve the quality of CMR reconstruction tasks. These findings provide valuable insights for  advancing DL approaches in the field of CMR reconstruction. 


\section*{Acknowledgment}
This publication has emanated from research conducted with the financial support of Science Foundation Ireland under Grant numbers 18/CRT/6183 and 12/RC/2289\_P2.





\bibliographystyle{IEEEtran}
\bibliography{IEEEabrv, midl-samplebibliography-short}




\vspace{12pt}

\end{document}